\documentclass[%
 reprint,
 amsmath,amssymb,
 aps,
superscriptaddress
]{revtex4-2}

\usepackage{graphicx}
\usepackage{dcolumn}
\usepackage{bm}
\usepackage[dvipsnames]{xcolor}

\usepackage{hyperref}

\begin{document}

\preprint{APS/123-QED}

\title{Forward-backward multiplicity and momentum correlations in $pp$ and $pPb$ collisions at the LHC energies}

\author{Joyati Mondal}
\email{joyati254@gmail.com}
\affiliation{Nuclear and Particle Physics Research Centre, Department of Physics,\\ Jadavpur University, Kolkata - 700032, India}

\author{Hirak Koley}
\affiliation{Nuclear and Particle Physics Research Centre, Department of Physics,\\ Jadavpur University, Kolkata - 700032, India}

\author{Somnath Kar}
\email{somnathkar11@gmail.com}
\affiliation{University Department of Physics, Kolhan University, Chaibasa - 833201, India}
\affiliation{Nuclear and Particle Physics Research Centre, Department of Physics,\\ Jadavpur University, Kolkata - 700032, India}

\author{Premomoy Ghosh}
\affiliation{Nuclear and Particle Physics Research Centre, Department of Physics,\\ Jadavpur University, Kolkata - 700032, India}

\author{Argha Deb}
\affiliation{Nuclear and Particle Physics Research Centre, Department of Physics,\\ Jadavpur University, Kolkata - 700032, India}
\affiliation{School of Studies in Environmental Radiation \&\\ Archaeological Sciences, Jadavpur University, Kolkata - 700032, India}

\author{Mitali Mondal}
\email{mitalimon@gmail.com}
\affiliation{Nuclear and Particle Physics Research Centre, Department of Physics,\\ Jadavpur University, Kolkata - 700032, India}
\affiliation{School of Studies in Environmental Radiation \&\\ Archaeological Sciences, Jadavpur University, Kolkata - 700032, India}

%

%
%
%
%

\date{\today}

\begin{abstract}
Correlations and fluctuations between produced particles in an ultra-relativistic nuclear collision remain one of the successor to understand the basics of the particle production mechanism. More differential tools like Forward-Backward (FB) correlations between particles from two different phase-space further strengthened our cognizance. We have studied the strength of FB correlations in terms of charged particle multiplicity and summed transverse momentum for proton-proton ($pp$) and proton-lead ($pPb$) collisions at the centre-of-mass energies $\sqrt{s}$ = 13 TeV and $\sqrt{s_{\rm NN}}$ = 5.02 TeV respectively for the EPOS3 simulated events with hydrodynamical evolution of produced particles. Furthermore, the correlation strengths are separately obtained for the particles coming from the core and the corona. 
FB correlation strengths are examined as a function of psedorapidity gap ($\eta_{gap}$), psedorapidity window-width ($\delta\eta$), centre-of-mass energy ($\sqrt{s}$), minimum transverse momentum ($p_{Tmin}$) and different multiplicity classes following standard kinematical cuts used by the ALICE and the ATLAS experiments at the LHC for all three EPOS3 event samples. EPOS3 model shows a similar trend of FB multiplicity and momentum correlation strengths for both $pp$ \& $pPb$ systems, though the correlation strengths are found to be larger for $pPb$ system than $pp$ system. Moreover, $\delta\eta$-weighted average of FB correlation strengths as a function of  different center-of-mass energies for $pp$ collisions delineates a tendency of saturation at very high energies. 
\end{abstract}
\maketitle


\section{Introduction}
The formation of a hot dense medium of quasi-free quarks and gluons, known as the Quark-Gluon Plasma, in relativistic heavy-ion collisions at the Relativistic Heavy Ion Collider (RHIC) and at the Large Hadron Collider (LHC) unlatch the path to understand the early Universe as well as to testify the theory of the strong interactions between quarks mediated by gluons~\cite{intro1, intro2, intro3}. The relativistic viscous hydrodynamic calculations~\cite{intro4, intro5} have been found to be most successful in explaining the properties of the produced hot and dense matter in heavy-ion collisions and demonstrate the space-time evolution of the medium through observables such as harmonic flow ($v_{n}$)~\cite{intro6, intro7, intro8} which represent the translation of initial state spatial inhomogeneities to the final state momentum anisotropies.

In heavy-ion collisions, the initial energy density fluctuates strongly event to event which leads to the fluctuations of the space-time evolution of the produced medium in the final state. Owing to the viscous hydrodynamics, such density fluctuations are manifested as anisotropic harmonic flow. The large initial state fluctuations effectuate the observed long-range correlations (LRC) between final state particles which are observed as correlation between multiplicity densities in different pseudorapidity ($\eta$)-windows~\cite{intro9}. Another aspect of longitudinal multiplicity correlations is the short-range correlations (SRC), localized over a smaller range of $\eta$, manifested in the single-jet, mini-jets, resonance decays etc. 
Forward-Backward (FB) correlations between charged particle multiplicities or transverse momenta in two symmetrically separated $\eta$-windows about the collision vertex promote us to differentiate between LRC and SRC components~\cite{JM1} and to study the dynamics of particle production mechanism in high-energy hadron or nuclear collisions.

Positive FB multiplicity correlation strength was first observed in $p\bar{p}$ collisions at $\sqrt{s}$ = 540 GeV at CERN SPS collider~\cite{JM4}
and it has been then rigorously inspected in $p\bar{p}$ collisions at ISR energies from $\sqrt{s}$ = 200 GeV to 900 GeV~\cite{JM5, JM6}. Later, FB correlations were also examined in $pp$ and $p\bar{p}$ collisions over wider range of collision energies~\cite{JM7, JM8, JM9}. No significant FB multiplicity correlation has been reported in $e^{+}e^{-}$ collisions~\cite{JM10} whereas a very weak correlation strength was observed in $e^{+}e^{-}$ annihilation~\cite{JM11, JM12}. CLAN structure  had been used for better understanding of observed stronger positive FB correlation in $pp$ and $p\bar{p}$ collisions compared to weak correlation in $e^{+}e^{-}$ annihilation~\cite{JM13}. There are also significant positive FB correlation values reported in different collision systems e.g. $pp$, $pA$ and $AA$ collisions~\cite{intro9, JM14, JM15, JM16, JM17}.  The ATLAS~\cite{JM16} and the ALICE~\cite{JM17} collaborations at the LHC reported strong FB correlations in $pp$ collisions at $\sqrt{s}$ = 0.9, 2.76 and 7 TeV which contradicts STAR collaboration findings of weak correlation~\cite{JM15}. 

To explain experimental results, numerous theoretical models have been put forth.  In Dual Parton Model (DPM)~\cite{JM2} a Pomeron exchange between colliding hadrons was initially thought of as an inelastic scattering, later, the idea of many Pomeron exchanges was implemented. DPM projected that particles created in two well selected rapidity intervals would have a significant long-range correlation~\cite{DPM2}. The Quark Gluon String Model (QGSM)~\cite{QGSM1, QGSM2} is similar to DPM with some essential differences. In this model new objects, quark-gluon strings are formed which fragment into hadrons and resonances. The QGSM model successfully described ALICE data and concluded that the multistring processes due to multi-Pomeron exchanges were the main contributor to the FB correlations. The String Fusion Model (SFM)~\cite{SFMMC1, SFMMC2} incorporates the string fusion phenomenon and is based on the Parton model of strong interactions. The SFM framework is used to investigate the characteristics of the strongly intensive variable that characterizes correlations between the number of particles in two separated rapidity intervals in $pp$ interactions at LHC energy~\cite{SFM1}. Using various dynamics of the string interaction assumptions, correlations between multiplicities and average transverse momentum are accomplished in the percolating colour strings picture~\cite{StringPer1, StringPer2}. A string percolation process in $pp$ collisions was used to study the FB correlations~\cite{StringPer1} and observed an approximately constant FB correlation over a substantial range of rapidity window. 

The correlations between mean transverse momentum and multiplicity of charged particles in $pp$ and $p\bar{p}$ collisions at $\sqrt{s}$ from 17 GeV to 7 TeV are studied using a modified multi-pomeron exchange model in which string collectivity has been included in an effective way~\cite{MultPomNew}. In $pp$, $pPb$, and $PbPb$ collisions at LHC energies it is explored using a dipole based Monte Carlo String Fusion Model~\cite{SFMMC3}. According to the Color Glass Condensate model~\cite{CGC1, CGC2, CGC3} long-range rapidity correlations continue throughout the development of the quark–gluon plasma that result from the collision. Using a model that regards strings as independent identical emitters, the forward-backward (FB) charged particle multiplicity correlations between windows spaced apart in rapidity and azimuth are investigated in Ref.~\cite{Vechernin1}. Theoretical background of long-range correlations in heavy-ion collisions has been studied using a Monte-Carlo method in Ref.~\cite{Vechernin2}.

The high multiplicity data of $pp$ and $pPb$ collisions at the LHC and $dAu$ collisions at the RHIC show some collective-like features resembling the heavy-ion collision~\cite{intro15, intro16, intro17, intro18, intro19, intro20, intro20N2, intro20N4}. The two-particle correlation studies in high multiplicity $pp/pPb$ collisions showed the heavy-ion signature; ``the ridge" which triggered many theoretical discussion/interpretations on the origin of it. Recent studies~\cite{intro20ERidge, intro20ERidge1} show that the hydrodynamical modelling which remains successful in explaining many features of heavy-ion collisions, is also found to be applicable in small collision systems. We discussed in our previous article~\cite{intro14} on how EPOS3 model~\cite{epos1} with hydrodynamical evolution of produced particles (referred as ``with hydro" in rest of the texts) successfully reproduced many features of small collision systems at the LHC energies~\cite{intro20a}. Furthermore, we investigated the FB multiplicity and momentum correlations using EPOS3 model by switching ON/OFF hydrodynamical evolution of produced particles which does not affect much the final outcomes. Studies using different models show that the FB correlation strength is found to be decreasing with increasing nuclear size upon the selected $\eta$-windows and with increasing collision energy for a fixed collision system~\cite{intro21, intro22}. It has also been proposed that instead of the contribution coming from particle production in initial stages of collisions, subsequent stage could modify the behaviour of FB correlations and hadron nucleus collision is expected to give more information on the whole scenario. Keeping this in mind, a comparative analysis of $pp$ and $pPb$ systems has been performed to improve our current understanding of FB phenomena.
We have inspected FB multiplicity and momentum correlation in $pp$ and $pPb$ collisions at the centre-of-mass energy $\sqrt{s}$ = 13 TeV and $\sqrt{s_{\rm NN}}$ = 5.02 TeV respectively using EPOS3 simulated with hydro events. In order to clarify the functions of each component of the model that contribute to the outcomes, we further divided the model into core and corona approaches. The energy density of the strings in the core is sufficient to activate the hydrodynamically evolving QGP description. In the corona, hadron creation from nucleon-nucleon collisions is viewed as an independent phenomenon~\cite{epos1}.

The rest of the paper is organized as follows: the definition of FB multiplicity and momentum correlation coefficients are introduced in Section~\ref{sec2}. Section~\ref{sec3} describes briefly the basic principles of EPOS3 model and sample size of generated events. Choice of EPOS3 simulated events and FB windows are illustrated in Section~\ref{sec4}. In Section~\ref{sec5}, the dependence of FB correlation strength by varying $\eta_{gap}$, $\delta\eta$, $p_{T_{min}}$ and different multiplicity classes have been discussed in details. More importantly, the behaviour of $\delta\eta$-weighted average of FB multiplicity and momentum correlation strengths as a function of centre-of-mass energy using EPOS3 simulated $pp$ events have been studied for the first time. Finally, conclusions are drawn in Section~\ref{sec6}.
\section{Forward-backward Charged Particle Multiplicity and Momentum Correlation Coefficient}\label{sec2}
Forward-backward correlations are measured between different observables in separated $\eta$-intervals, namely n--n (the correlation between charged-particle multiplicities), $p_{T}$--$p_{T}$ (the correlation between mean or summed transverse momenta of charged particles) and $p_{T}$--n (the correlation between mean or summed transverse momenta in one pseudorapidity interval and the multiplicity of charged particles in another pseudorapidity interval)~\cite{2ndSec1}. Two $\eta$-intervals, one from the forward and another from the backward window, are symmetrically chosen around the collision centre. The detailed window construction has already been shown and discussed in Ref.~\cite{intro14}.

A linear relationship between average charged particle multiplicity in backward window ($\langle N_b\rangle_{N_f}$) and the charged particle multiplicity in the forward window ($N_f$) has been reported and discussed in Ref.~\cite{JM5, JM6}:
\begin{equation}
\langle N_{b}\rangle_{N_{f}} = a + b_{corr} (mult) N_{f}
\label{eq1}
\end{equation}
Here, FB multiplicity correlation strength is characterized by $b_{corr}(mult)$. 
Considering the linear relation between ($\langle N_b\rangle_{N_f}$) and $N_f$, $b_{corr}(mult)$ can be determined using the following Pearson Correlation Coefficient:
\begin{equation}
b_{corr} (mult) = \frac{\langle N_{f} N_{b}\rangle - \langle N_{f}\rangle \langle N_{b}\rangle}{\langle N_{f}^{2}\rangle - \langle N_{f}\rangle^{2}} 
\label{eq2}
\end{equation}
The measurement of FB multiplicity correlation coefficient $b_{corr}(mult)$ is defiled by the so-called ``volume fluctuations", which arises due to the event-by-event fluctuations of the number of the participating nucleons~\cite{JM41Sec2, JM41Sec3}. Hence, we have considered an intensive observable like the sum of the absolute transverse momentum of charged-particles within the selected $\eta$ windows to reduce the contribution of volume fluctuations. Similar to the multiplicity correlation we have estimated FB momentum correlation coefficient, $b_{corr}(\Sigma p_T)$ using the following formula: 
\begin{equation}
\small b_{corr} (\Sigma p_{T}) = \frac{\langle \Sigma p_{T_{f}} \Sigma p_{T_{b}}\rangle -  \langle \Sigma p_{T_{f}}\rangle \langle\Sigma p_{T_{b}}\rangle}{\langle(\Sigma p_{T_{f}})^{2}\rangle -  \langle\Sigma p_{T_{f}}\rangle^{2}}
\label{eq3}
\end{equation}
Here, $\Sigma p_{T_{f}}$ ($\Sigma p_{T_{b}}$) denotes the event averaged transverse momenta of charged-particles in forward (backward) window.

An intuitive observable, $\delta\eta$-weighted average of FB multiplicity  and momentum correlation strength has been introduced for the first time and is defined as follows:
\begin{equation}
\small <b_{corr} (mult/\Sigma p_{T})>_{\delta\eta} = \frac{\Sigma_{i} b_{corr} (mult/\Sigma p_{T})_{i} \delta\eta_{i}}{\Sigma_{i} \delta\eta_{i}}
\label{eq4}
\end{equation}
The behaviour of such observable has been studied as a function of the centre-of-mass energy in $pp$ collisions taking into account of our earlier measurements in smaller collision system~\cite{intro14}. 
\begin{figure}
\centering
\includegraphics[scale=0.28,keepaspectratio]{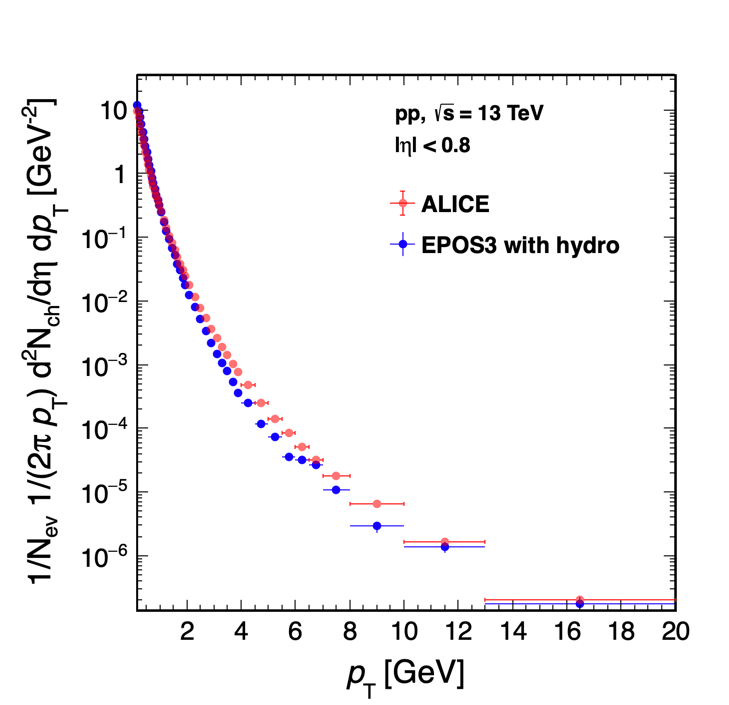}
\includegraphics[scale=0.29,keepaspectratio]{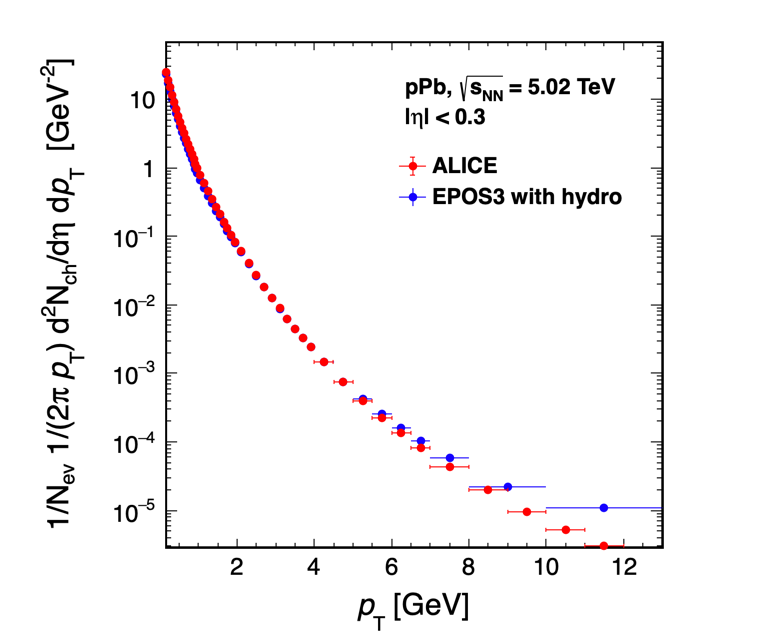}
\caption{Charged-particle invariant yields as a function of $p_{\rm T}$ in $pp$ collisions at $\sqrt{s}$ = 13 TeV (top) and in $pPb$ collisions at $\sqrt{s_{\rm NN}}$ = 5.02 TeV (bottom) compared to ALICE data~\cite{aliceEpos3pp, aliceEpos3pPb}.}
\label{figPT}
\end{figure}

\begin{figure}
\centering
\includegraphics[scale=0.30,keepaspectratio]{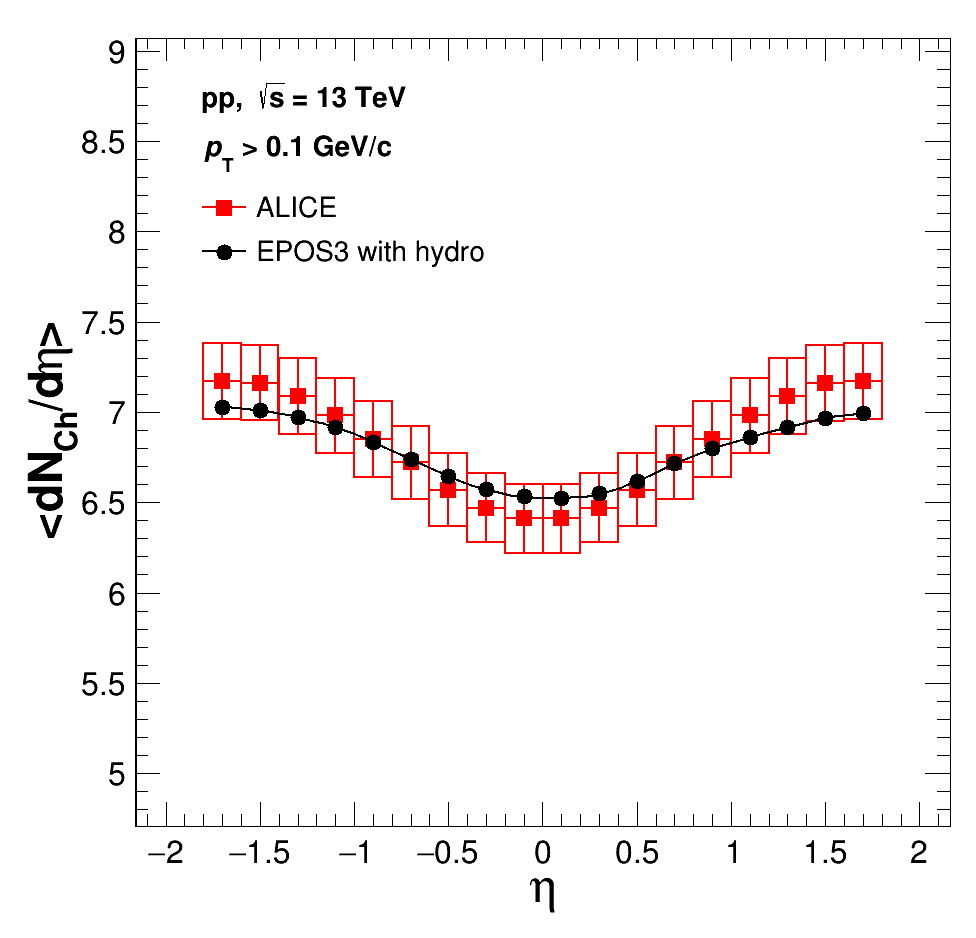}
\includegraphics[scale=0.30,keepaspectratio]{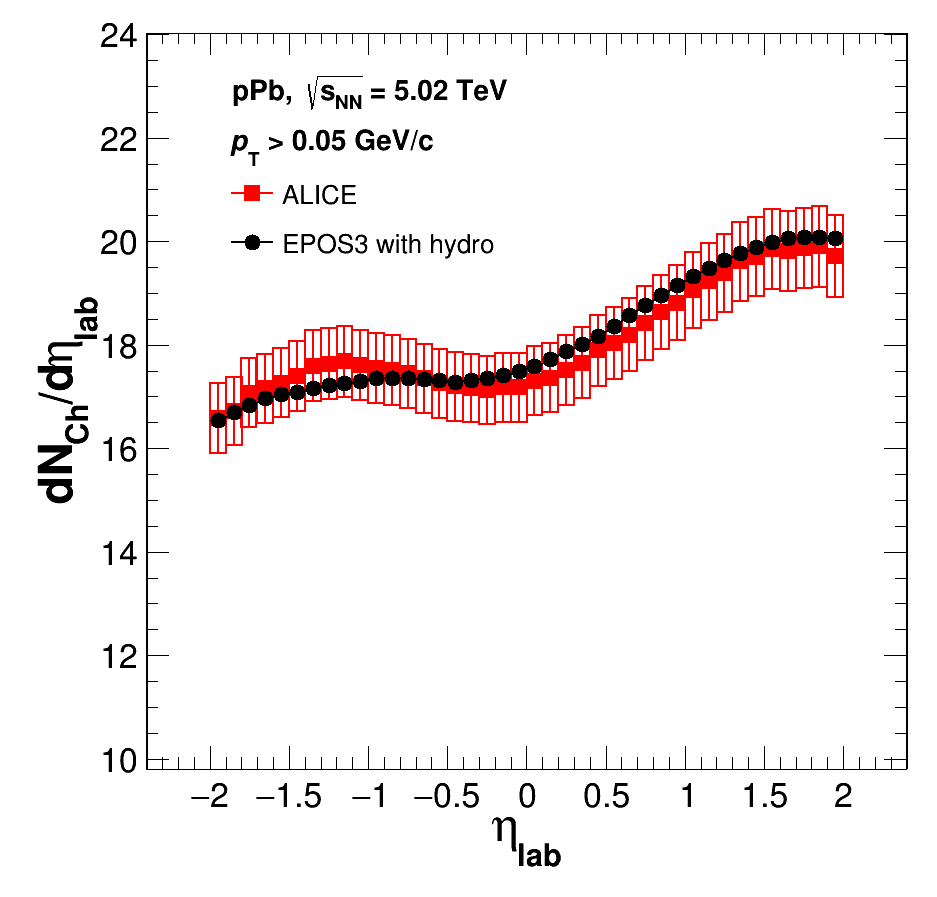}
\caption{(Average) Pseudorapidity density of charged particles in ($pp$ collisions at $\sqrt{s}$ = 13 TeV (top)) $pPb$ collisions at $\sqrt{s_{\rm NN}}$ = 5.02 TeV (bottom) compared to ALICE data~\cite{aliceEpos3pp, AlicepPbEta}.}
\label{figEta}
\end{figure}
\section{EPOS3 Model}\label{sec3}
The EPOS3 model~\cite{epos1} is based on Gribov-Regge multiple scattering theory. In this approach an individual scattering is labeled as a ‘Pomeron’. A pomeron creates a parton ladder which may be considered as longitudinal flux tube carrying the transverse momentum from the initial hard scatterings~\cite{epos2}. In a collision, many elementary parton-parton hard scatterings form a large number of flux tubes that expand and are fragmented into string segments. Higher string density forms the so-called ``core" which undergoes a three-dimensional (3D)+1 viscous hydrodynamical evolution expecting no jet parton escape and hadronizes via usual Cooper-Frye formalism at a ``hadronization temperature", T$_{\rm H}$.
Another part of lower string density forms the so-called ``corona" where we can expect the escape of jet partons. Such string segments having high transverse momentum that are close to the surface leave the bulk matter and hadronize (including jet hadrons) via the Schwinger mechanism. The phase transition from parton to hadron follows a realistic equation of state which is compatible with the lattice gauge results with subsequent hadronic cascade using UrQMD model~\cite{urqmd}.

Using EPOS3 model, we have generated 3 million minimum-bias $pp$ events at $\sqrt{s}$ = 13 TeV and $pPb$ events at $\sqrt{s_{\rm NN}}$ = 5.02 TeV. On top of the minimum bias analysis, a more differential approach has been introduced by taking particles coming from either core or corona and we have varied certain model parameter in order to achieve it. We have measured FB multiplicity and momentum correlations for the EPOS3 generated events with all charged particles and particles from core and corona. 

To validate the generated event samples of different centre-of-mass energies, we have compared minimum bias EPOS3 simulated events with ALICE data~\cite{aliceEpos3pp, aliceEpos3pPb, AlicepPbEta}. Fig.~\ref{figPT} shows that the invariant yields of  charged-particles as a function of $p_{\rm T}$ as measured by ALICE experiment in $pp$ collisions at $\sqrt{s}$ =  13 TeV (top) and in $pPb$ collisions at $\sqrt{s_{\rm NN}}$ = 5.02 TeV (bottom) have been successfully reproduced by the EPOS3 simulated events at the chosen energies. Average pseudorapidity density and pseudorapidity density of charged-particles has been plotted in Fig.~\ref{figEta} for EPOS3 simulated $pp$ events at $\sqrt{s}$ = 13 TeV (top) and pPb events at $\sqrt{s_{\rm NN}}$  =  5.02 TeV (bottom) respectively. 
Compared results reflect that EPOS3 simulated events agreed well with the experimental measurements in the chosen kinematic intervals~\cite{aliceEpos3pp, AlicepPbEta}.
\section{Events \& FB Window Selection}\label{sec4}
Events are selected with a minimum of two charged particles in the chosen kinematic interval. 
The whole analyses for the $pp$ and $pPb$ events have been carried out following ALICE~\cite{JM17} and ATLAS~\cite{JM16} kinematics. By the word ALICE kinematics we mean the cuts on the kinematic variables $p_{\rm T}$ and $\eta$ as 0.3 $< p_{\rm T} <$1.5 GeV/c and $|\eta| <$ 0.8 respectively. Similarly for the ATLAS kinematics we use $p_{\rm T}>$ 0.1 GeV/c  and $|\eta| <$ 2.5. Only caveat is those cuts were used for lower centre-of-mass energies for $pp$ collisions by the ALICE and the ATLAS Collaborations. 
\section{Results \& Discussions}\label{sec5}
\begin{figure}
\centering
\includegraphics[scale=0.32,keepaspectratio]{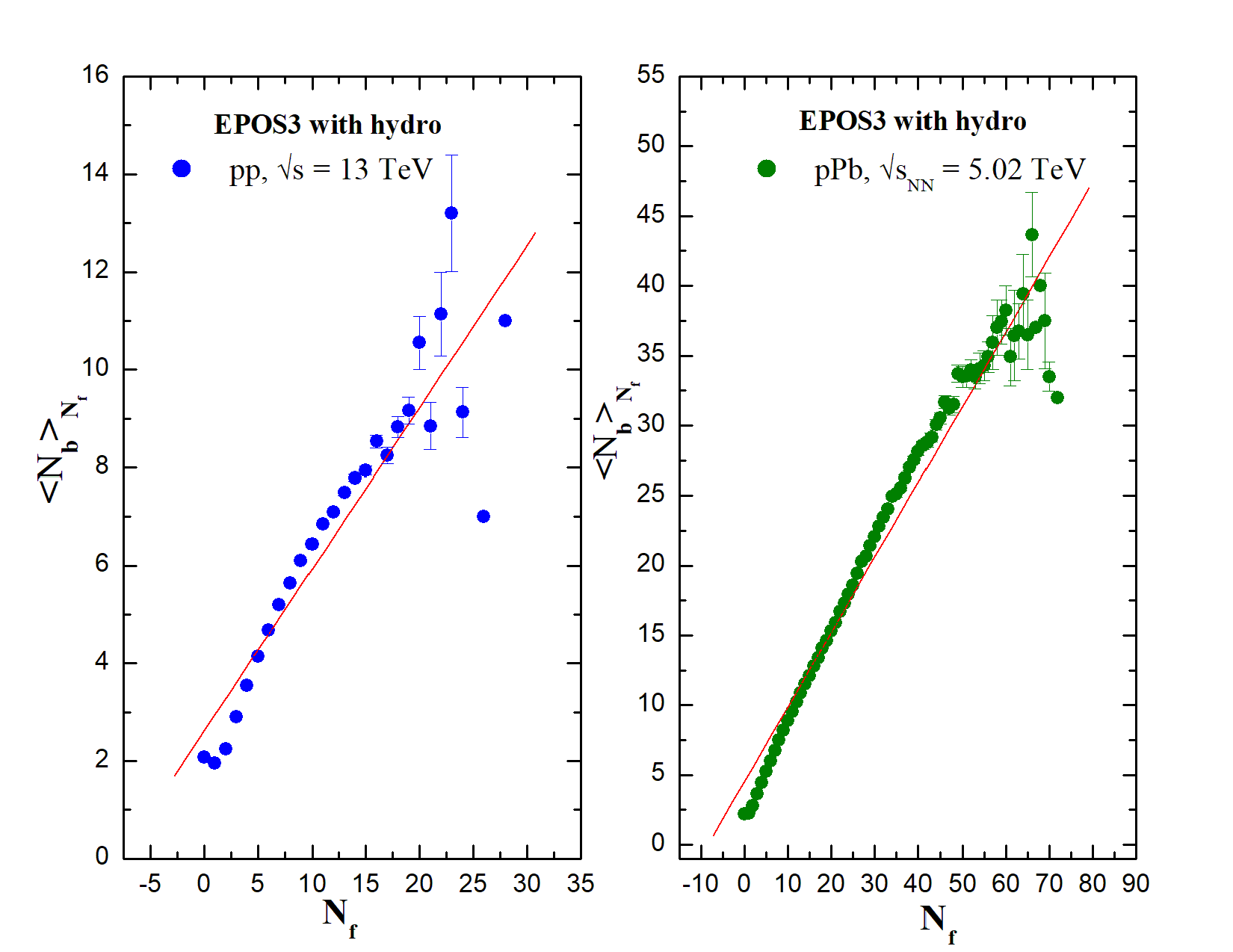}
\caption{Variation of $\langle N_{b}\rangle_{N_{f}}$ with $N_{f}$ for FB window width $\delta\eta$ = 0.6 and $\eta_{gap}$ = 0.4 for 
EPOS3 generated $pp$ events  at  $\sqrt{s}$ = 13 TeV (left panel) and $pPb$ events (right panel) at  $\sqrt{s_{\rm NN}}$ = 5.02 TeV. A linear fit has been performed (red line) for both the systems.}
\label{figNbNf}
\end{figure}

We have calculated and plotted in Fig.~\ref{figNbNf} the average backward multiplicity ($\langle N_{b}\rangle_{N_{f}}$) for each fixed value of forward multiplicity $N_f$ for window width $\delta\eta =$ 0.6 and $\eta_{gap} =$ 0.4 for EPOS3 simulated $pp$ events at $\sqrt{s}$ = 13 TeV (left panel) and $pPb$ events at $\sqrt{s_{\rm NN}}$ = 5.02 TeV (right panel). From the scatter plots we can see a linear relationship between $\langle N_{b}\rangle_{N_{f}}$ and $N_f$. A linear fit has been displayed by the red lines in both panels. Slope of these lines actually quantify the correlation strength between multiplicities in F $\&$ B windows. 
We have, therefore, applied Pearson Correlation Coefficient formula described in Eq.~\ref{eq2} to compute FB multiplicity correlation strengths. To eliminate the incorporated volume fluctuations in FB multiplicity correlation, we have evaluated FB momentum correlation coefficient, $b_{corr} (\Sigma p_{\rm T} )$  using Eq.~\ref{eq3} for the EPOS3 simulated $pp$ events at $\sqrt{s}$ = 13 TeV and $pPb$ events at $\sqrt{s_{\rm NN}}$ = 5.02 TeV. 

\subsection{\bf Dependence on the gap between\\ FB windows ($\eta_{gap}$)}
\label{FBCorrEtagap}
\begin{figure}[hbt!]
\centering
\includegraphics[scale=0.30,keepaspectratio]{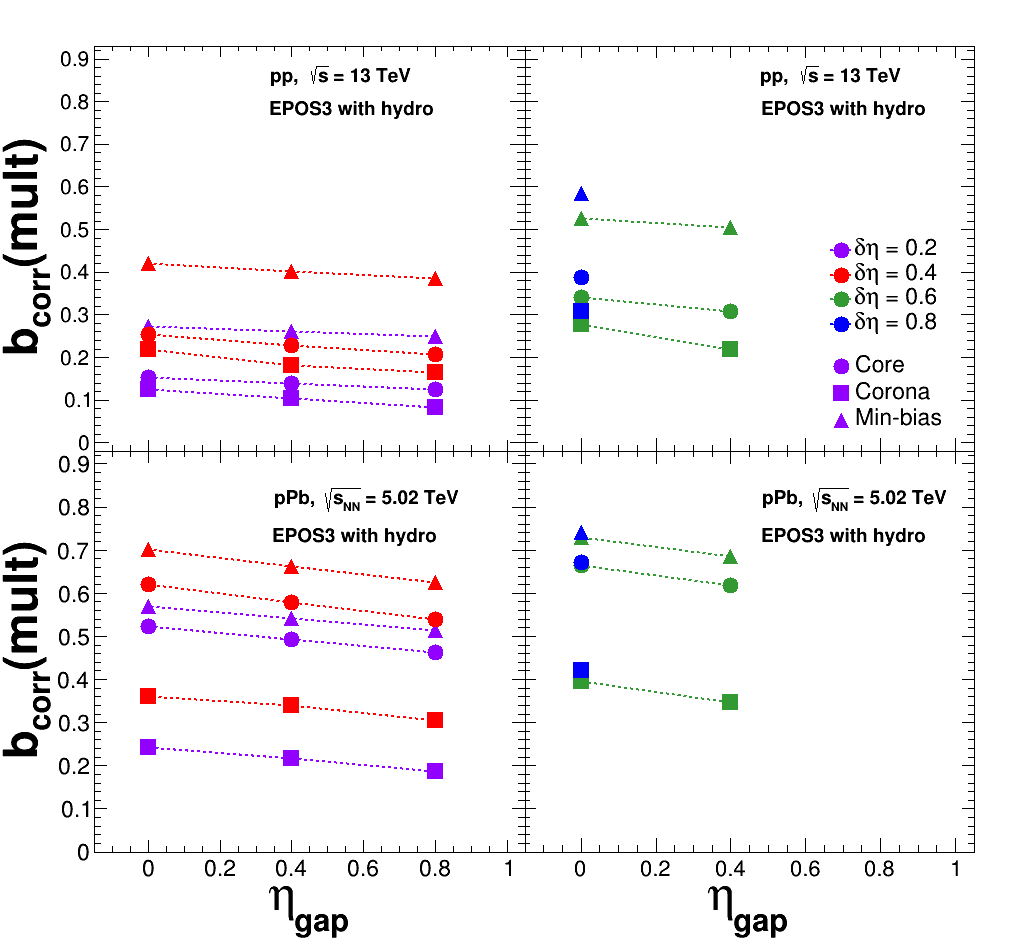}
\caption{FB multiplicity correlation strength, $b_{corr}$(mult) as a function of $\eta_{gap}$ for
$\delta\eta$ = 0.2, 0.4, 0.6 and 0.8 for EPOS3 generated all charged particles and particles from core \& corona in $pp$ collisions at $\sqrt{s}$ = 13 TeV (top panel) and $pPb$ collisions (bottom panel) at  $\sqrt{s_{\rm NN}}$ = 5.02 TeV. }
\label{bcorrMultEtaGap}
\end{figure}

\begin{figure}[hbt!]
\centering
\includegraphics[scale=0.28,keepaspectratio]{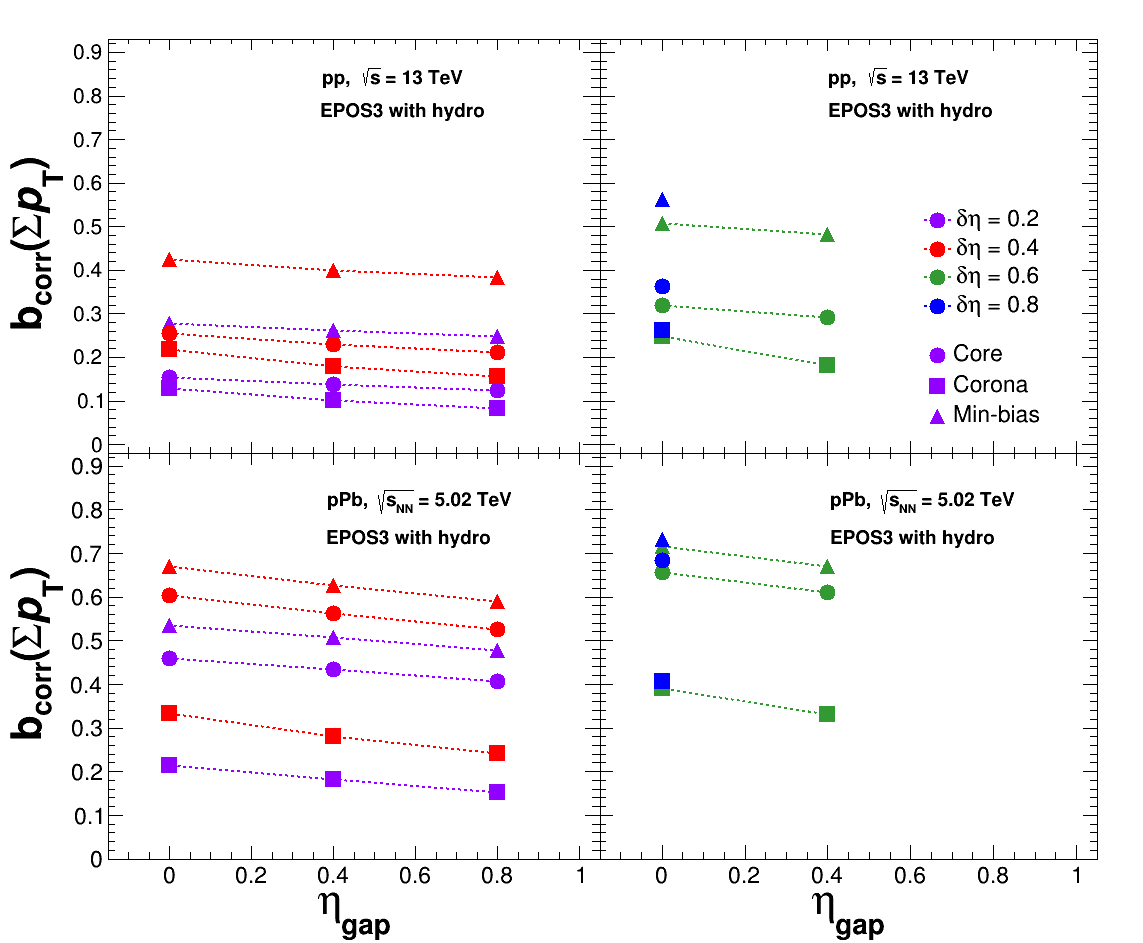}
\caption{Forward-backward summed-$p_{\rm T}$ correlation as a function of $\eta_{gap}$ for four window widths $\delta\eta$ 
= 0.2, 0.4, 0.6 and 0.8 for EPOS3 generated all charged particles and particles from core \& corona in $pp$ and $pPb$ collisions at $\sqrt{s}$ = 13 TeV (top panel) and $\sqrt{s_{\rm NN}}$ = 5.02 TeV (bottom panel) respectively.}
\label{bcorrSumPtEtaGap}
\end{figure}

The variation of FB multiplicity and momentum correlation coefficients with $\eta_{gap}$  for four different window widths ($\delta\eta$ = 0.2, 0.4, 0.6 \& 0.8) have been shown in Fig.~\ref{bcorrMultEtaGap} and Fig.~\ref{bcorrSumPtEtaGap} respectively for EPOS3 simulated all charged particles, only-core and only-corona particles in $pp$ collisions at $\sqrt{s}$ = 13 TeV (top panel) and $pPb$ collisions at $\sqrt{s_{\rm NN}}$ = 5.02 TeV (bottom panel). 
%
We have compared all three cases for window widths $\delta\eta$ = 0.2 \& 0.4 (left panel) and $\delta\eta$ = 0.6 \& 0.8 (right panel) for both $pp$ and $pPb$ events. We observed that for a fixed window width, the FB correlation strengths decrease slowly with the increasing $\eta_{gap}$ and increase with increasing $\delta\eta$ at a fixed $\eta_{gap}$ which resemble the trend at lower centre-of-mass energies in $pp$ collisions as described in our earlier study~\cite{intro14}. 

Quantitatively we found that the correlation strengths are larger for $pPb$ collisions than $pp$ collisions for all the chosen $\eta_{gap}$ and $\delta\eta$ combinations. The asymmetric nature of $pPb$ collisions where the proton collides with a nucleus having a larger number of sources compared to $pp$ collisions, results in a larger initial-state parton density in the lead nucleus compared to the proton. Such asymmetric collisions could have larger fluctuations in the final state which may contribute to stronger forward-backward correlation strength in $pPb$ collisions w.r.t $pp$ collisions. 

Interestingly, we have noticed that for $pPb$ events the correlation strengths decrease faster with increasing $\eta_{gap}$ as compared to $pp$ events. The SRC component depends strongly on collision system and it is asymmetric between forward and backward windows in $pPb$ collisions while the LRC component is nearly symmetric in all collision systems~\cite{intro9}. Thus, the faster dilution of SRC component at large $\eta_{gap}$ between forward and backward regions could be the reason behind the faster decrease of correlation strength in asymmetric $pPb$ collisions w.r.t symmetric $pp$ collisions. 

The dominance of correlation strength due to only-core particles is clearly visible over only-corona particles. Since the particles from corona are mostly dominated by jets or minijet partons, the paucity of LRC component results in smaller correlation strength for particles from corona over core at large $\eta_{\rm gap}$ for both the collision systems.

%
\subsection{\bf Dependence on the width of FB windows ($\delta\eta$)}
\begin{figure}
\centering
\includegraphics[scale=0.28,keepaspectratio]{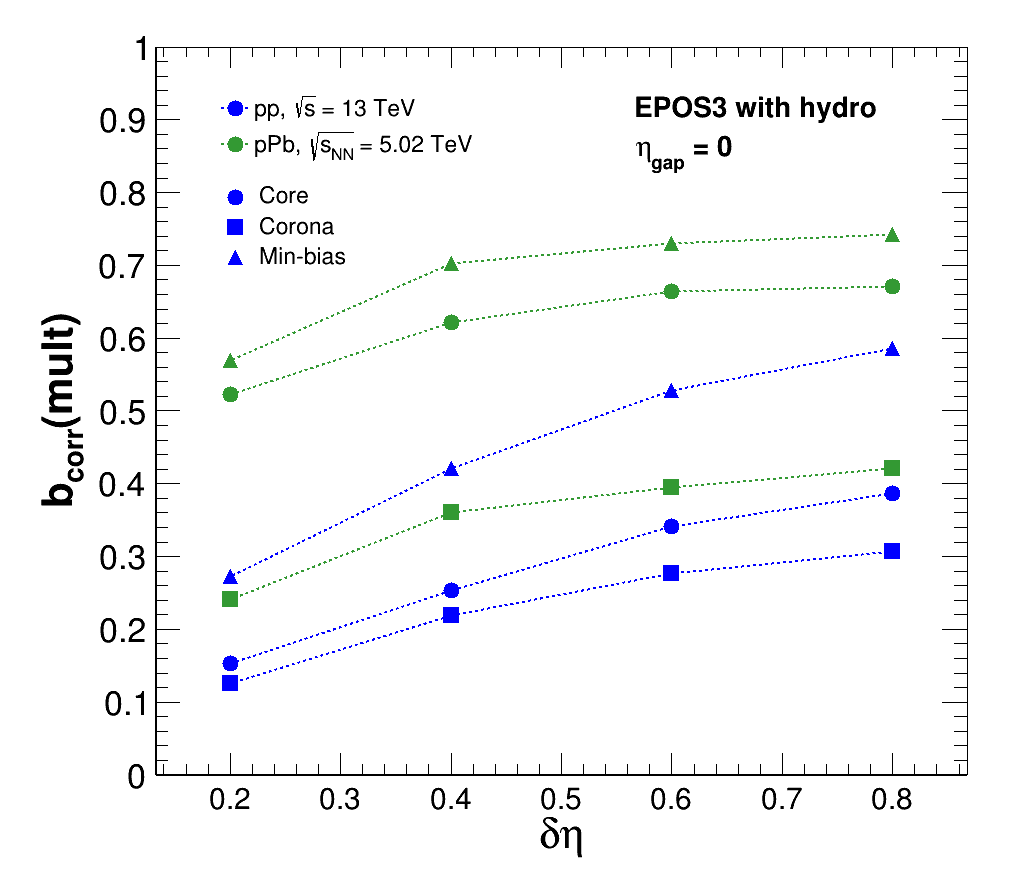}
\caption{FB multiplicity correlation strength, $b_{corr}$(mult) as a function of $\delta\eta$ for $\eta_{gap}$ = 0 using EPOS3 generated $pp$ and $pPb$ events at $\sqrt{s}$ = 13 TeV and $\sqrt{s_{\rm NN}}$ = 5.02 TeV respectively for all charged particles and particles from core \& corona.}
\label{figbCoMultDelEta}
\end{figure}

\begin{figure}[hbt!]
\centering
\includegraphics[scale=0.26,keepaspectratio]{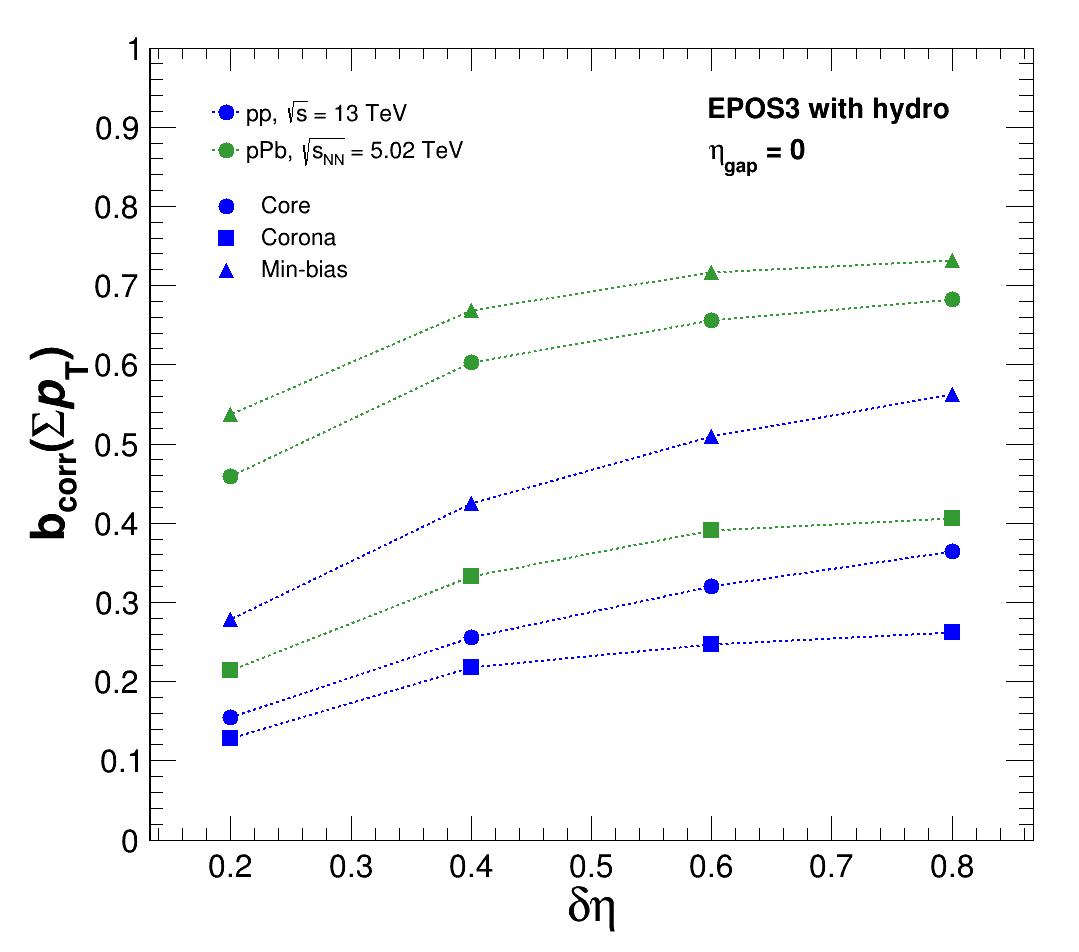}
\caption{Contributions of $b_{corr} (\Sigma p_{T})$ on $\delta\eta$ for $\eta_{gap}$ = 0 for EPOS3 generated $pp$ and $pPb$ events.}
\label{figbCoSumPtDelEta}
\end{figure}

The $\delta\eta$-dependence of FB multiplicity and momentum correlation coefficients for contiguous ($\eta_{gap}$ = 0), symmetrical windows with respect to collision centre are shown in Fig.~\ref{figbCoMultDelEta}  and Fig.~\ref{figbCoSumPtDelEta} for $\sqrt{s}$ = 13 TeV in $pp$ collisions and $\sqrt{s_{\rm NN}}$ =  5.02 TeV in $pPb$ collisions using EPOS3 simulated with hydro events. 
We have studied and presented the multiplicity and momentum correlation coefficients for the EPOS3 generated event samples with all charged particles and particles from core \& corona. We observed that the only-core and only-corona cases underestimate both the correlation strengths for minimum-bias event sample. The correlation coefficients increase non-linearly with $\delta\eta$ for both $pp$ and $pPb$ events though the values are higher in case of $pPb$ events which may be due to fact as described in~\ref{FBCorrEtagap}. The results are found to be similar to our earlier measurements qualitatively~\cite{intro14}, featuring the dominance of SRC component for the non-linear growth of the FB correlation strengths. As discussed and explained in ~\ref{FBCorrEtagap}, here also we have found that the correlation strengths w.r.t $\delta\eta$ are larger for only-core particles than only-corona particles.
\subsection{\bf Dependence on collision energy ($\sqrt{s}$)}
\begin{figure}[hbt!]
\centering
\includegraphics[scale=0.23,keepaspectratio]{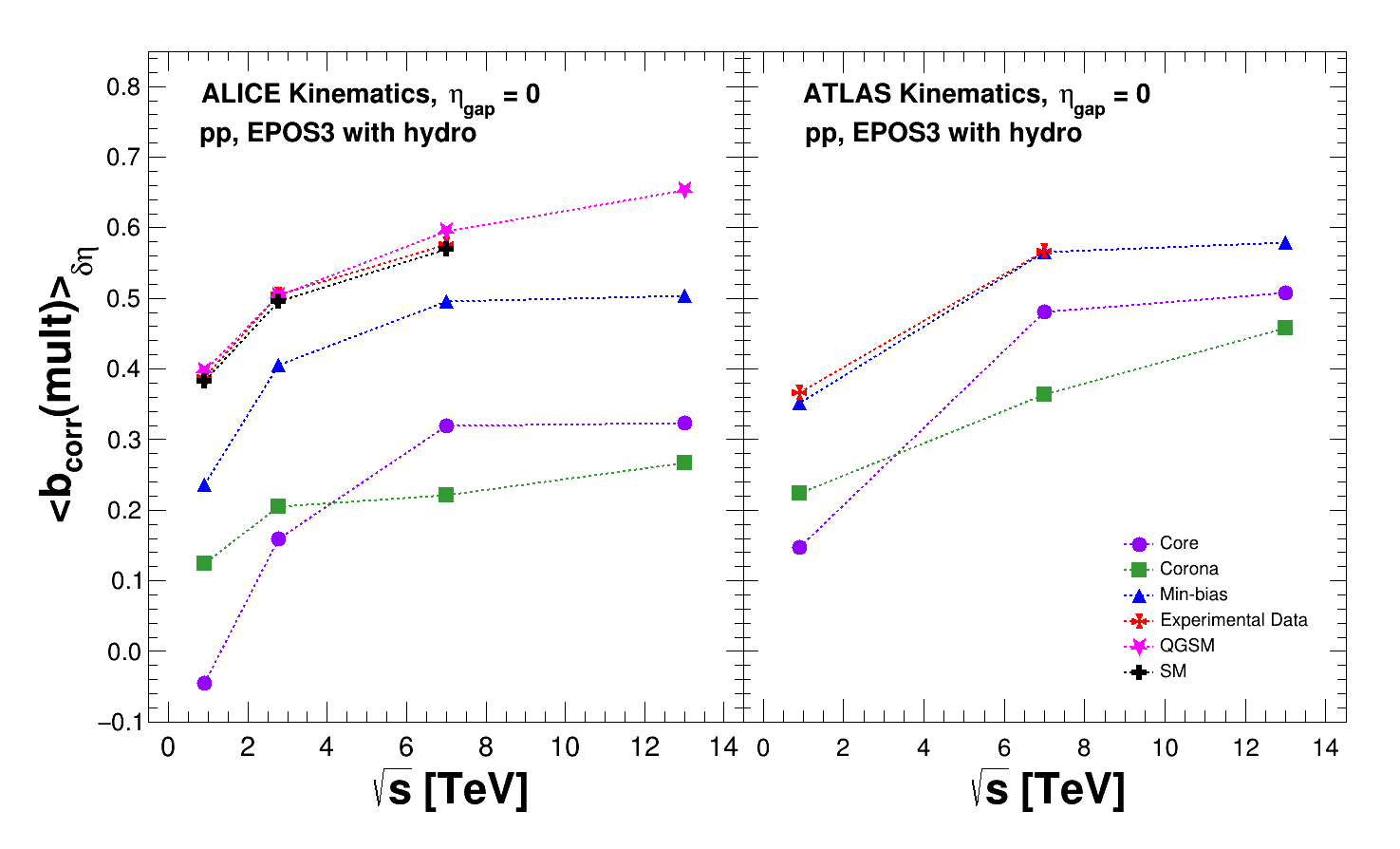}
\caption{Comparison of $\delta\eta$-weighted average FB multiplicity correlations, ($\langle b_{corr}(mult)\rangle_{\delta\eta}$) as a function of $\sqrt{s}$ for the EPOS3 simulated $pp$ events (all charged particles, core and corona) with derived ALICE (left), ATLAS (right) data and theoretical models (left).}
\label{fig5}
\end{figure}

\begin{figure}[hbt!]
\centering
\includegraphics[scale=0.23,keepaspectratio]{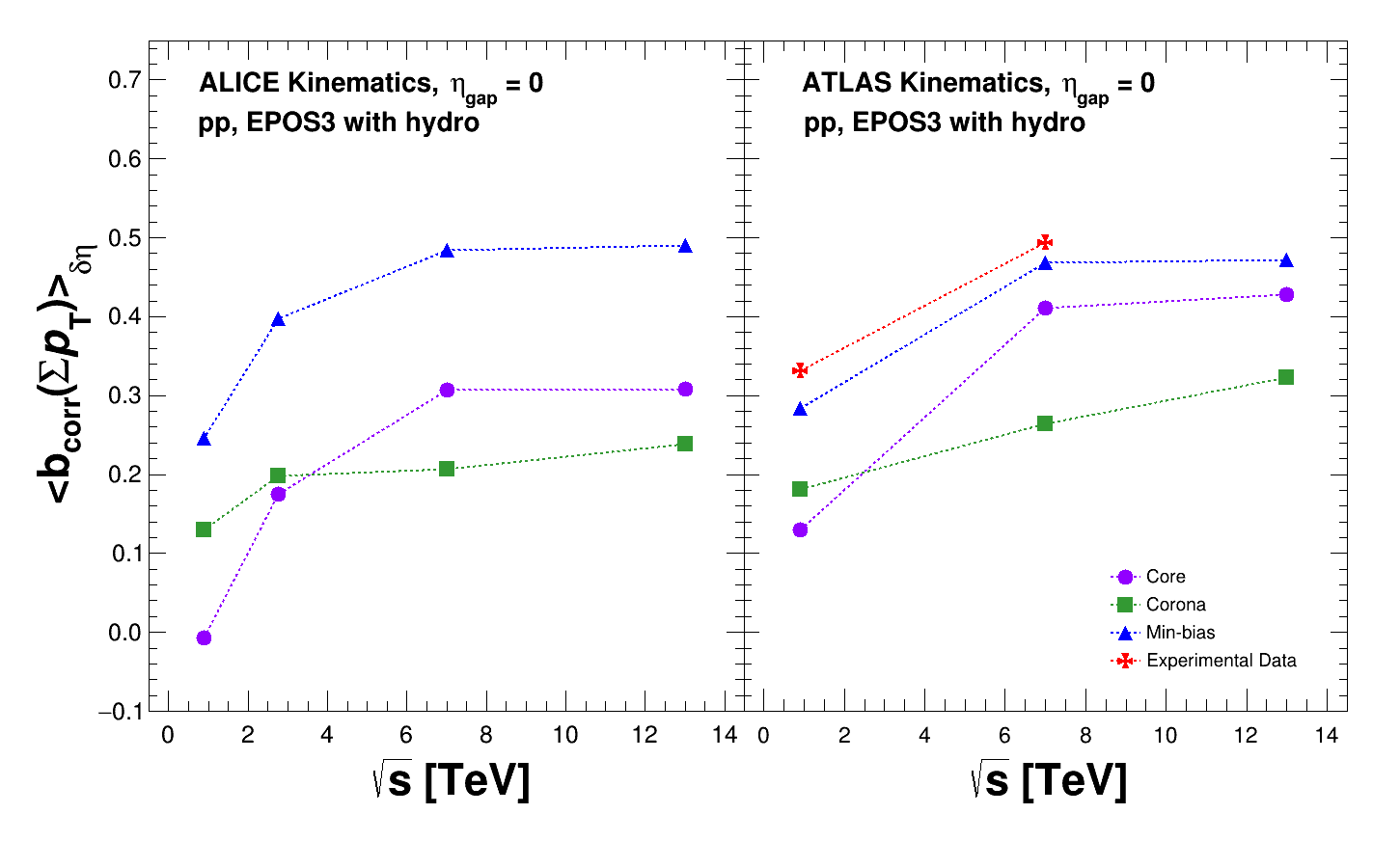}
\caption{Comparison of $\delta\eta$-weighted average FB summed-$p_{\rm T}$ correlations, ($\langle b_{corr}(\Sigma p_{\rm T})\rangle_{\delta\eta}$) as a function of $\sqrt{s}$ for EPOS3 simulated $pp$ events (all charged particles, core and corona) following ALICE kinematics (left) and with derived ATLAS (right) data.}
\label{fig9}
\end{figure}

We have examined the behaviour of $\delta\eta$-weighted average of FB multiplicity and momentum correlation strengths as a function of centre-of-mass energy using EPOS3 simulated $pp$ event samples. Such an unconventional measurement is still not available experimentally. Hence, to compare our findings in a systematic way we have evaluated $\delta\eta$-weighted average for the available experimental $b_{corr}(mult)$ and  $b_{corr}(\Sigma p_{\rm T})$ values for the ALICE~\cite{JM17} and the ATLAS~\cite{JM16} data. Fig.~\ref{fig5} and Fig.~\ref{fig9} show the $\delta\eta$-weighted average of FB multiplicity and momentum correlation as a function of centre-of-mass energy following ALICE and ATLAS kinematics. 

In the left panel of Fig.~\ref{fig5} and  Fig.~\ref{fig9}, we observed that initially $\langle b_{corr}(mult)\rangle_{\delta\eta}$ and $\langle b_{corr}(\Sigma p_{\rm T})\rangle_{\delta\eta}$ values increase rapidly with increasing $\sqrt{s}$ upto 2.76 TeV, then grows moderately upto $\sqrt{s}$ = 7 TeV for both EPOS3 simulated events and experimental data. For comparison, we have incorporated the results from other available theoretical models which also show  a similar trend for the $\delta\eta$-weighted average of FB correlations~\cite{QGSM2, Vechernin1}. 

It is also very interesting to find that for EPOS3 simulated events, the $\delta\eta$-weighted average of FB correlation strengths as a function of $\sqrt{s}$ lean towards saturation approximately beyond $\sqrt{s}$ = 7 TeV, where there is no available experimental data for such a measurement. The results from QGSM model although do not show such a strong saturation effect at higher energy. 

To gain a more nuanced understanding of the fascinating behavior observed in the $\delta\eta$-weighted average of FB correlation strengths, we computed this metric separately for the particles coming either from core or corona. We have found that the observed saturation at higher centre-of-mass energy predominantly due to the saturation for the only-core particles whereas the only-corona particles show increasing trend. 
In EPOS3 model, the corona is dominated by the high-$p_{T}$ particles whereas the core contains particles which undergo 3+1D hydro mimicking the formation of QGP-like medium~\cite{EPOS3CrCna}. 

The exchange of multiple pomerons between colliding particles~\cite{MultPom} in a collision remains the primary sources of fluctuations producing multiple particles in a correlated way. The multiplicity of produced particle and their transverse momentum are thus very much influenced by the initial conditions of a collision and in particluar are much more apparent in small collision systems like $pp$ or $pA$ where final state effects are less. In the CGC framework~\cite{CGC4, FGelis} it has been argued that at small $x \sim p_{T}/\sqrt{s}$, gluon density first grows and then gets saturated with increase in energy which results in moderate increase in charge particle multiplicity density in $pp$ or $pA$ collisions as the beam energy increases~\cite{multAlice}. Since the observable like b$_{corr}$ (mult or $\Sigma p_{\rm T}$) is an extensive quantity, it might show such saturation effect mainly because fluctuations associated with the number of sources get saturated~\cite{MultPom}.
Henceforth, the EPOS3 model based FB correlation analysis at higher center-of-mass energy encourages further experimental study. 
\subsection{\bf Dependence on the minimum transverse momentum ($p_{\rm T_{min}}$)}

\begin{figure}
\centering
\includegraphics[scale=0.30,keepaspectratio]{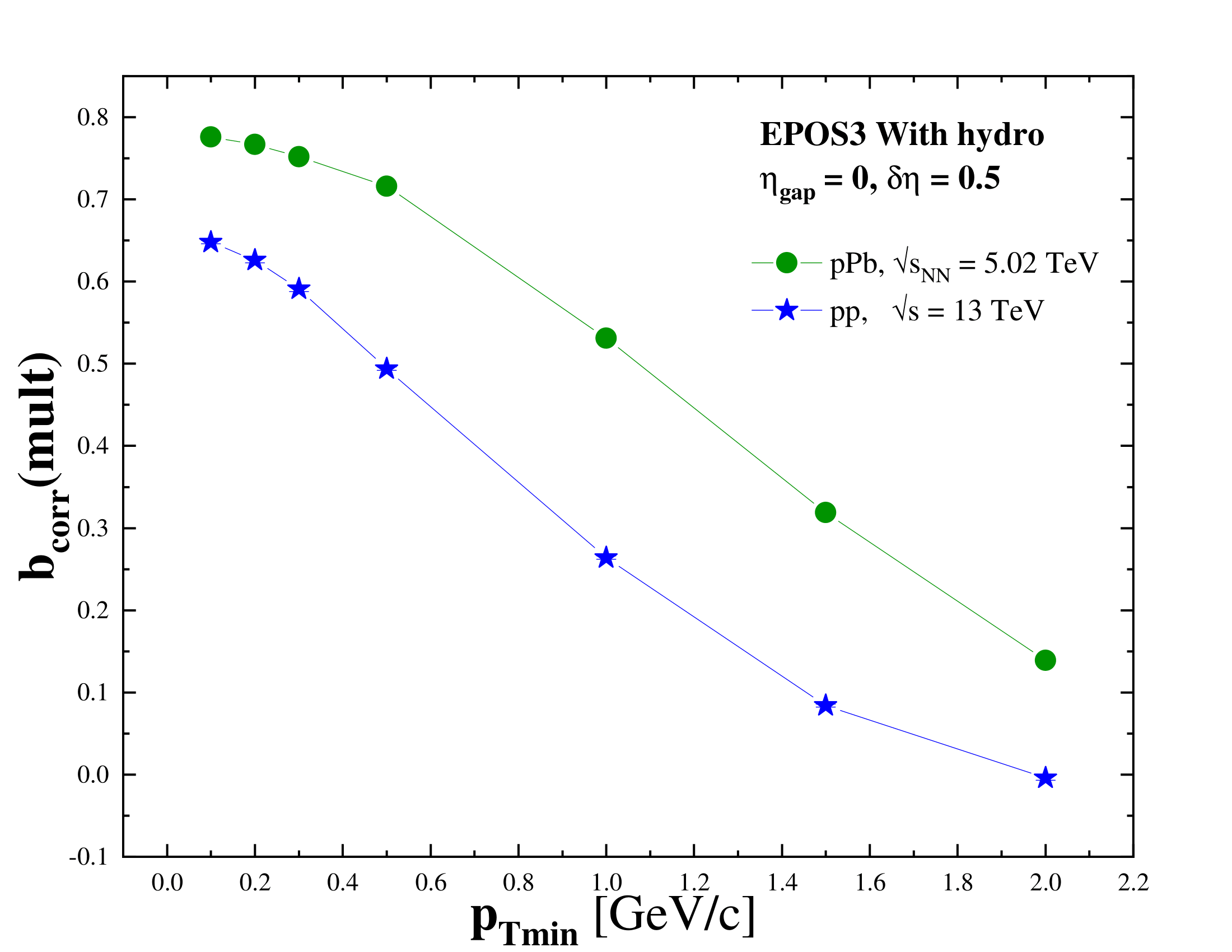}
\caption{Forward-backward multiplicity correlations as a function of $p_{\rm T_{min}}$ for window width $\delta\eta$ = 0.5 for EPOS3 simulated $pp$ and $pPb$ events.}
\label{fig6}
\end{figure}

\begin{figure}[hbt!]
\centering
\includegraphics[scale=0.30,keepaspectratio]{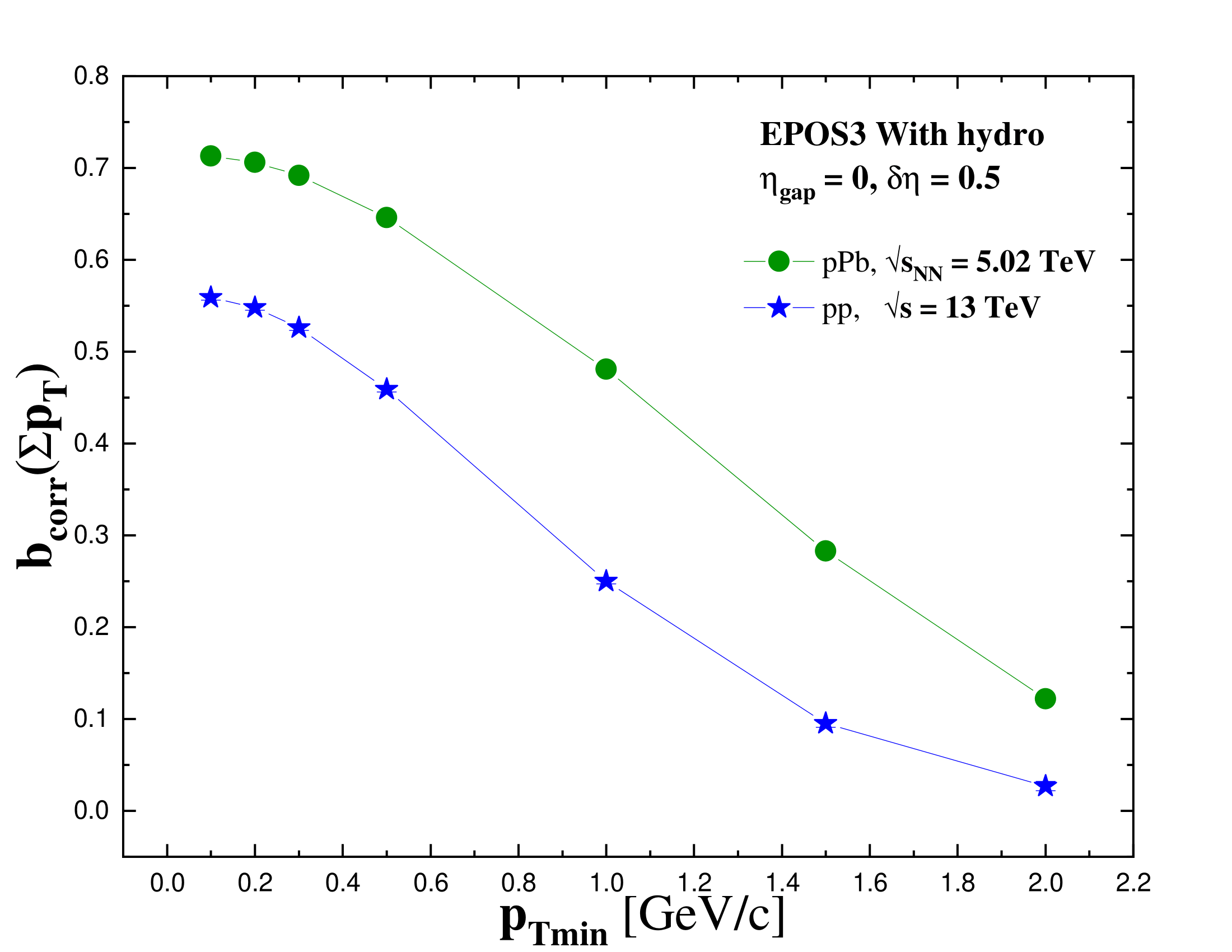}
\caption{Forward-backward summed-$p_{\rm T}$ correlation as a function of $p_{\rm T_{min}}$  for window width $\delta\eta$ = 0.5 for EPOS3 generated $pp$ and $pPb$ events.}
\label{fig10}
\end{figure}

The variation of FB multiplicity and momentum correlations with minimum transverse momentum of charged particles ($p_{\rm T_{min}}$) have been shown in Fig.~\ref{fig6} and Fig.~\ref{fig10} for both EPOS3 generated with hydro $pp$ events at $\sqrt{s}$ = 13 TeV and $pPb$ events at $\sqrt{s_{\rm NN}}$ = 5.02 TeV following ATLAS kinematics~\cite{JM16}.  We calculated the values of $b_{corr}$ at seven different levels of minimum transverse momentum ($p_{\rm T_{min}}$), specifically at $p_{\rm T_{min}}$ = 0.1, 0.2, 0.3, 0.5, 1.0, 1.5, and 2.0 GeV. These calculations were performed for symmetric forward-backward (FB) windows without any separation.
The multiplicity and momentum correlations strengths decrease rapidly with the increase of  $p_{\rm T_{min}}$ values for both $pp$ and $pPb$ events confirming similar trend at lower center-of-mass energies~\cite{intro14}. With the increase of $p_{\rm T_{min}}$, domination of LRC component decreases resulting weaker FB multiplicity correlation strength suggesting the transition from soft process to hard processes with increasing transverse momentum of the produced particles. As discussed in ~\ref{FBCorrEtagap}, here also the correlation strengths are found to be greater for $pPb$ collisions than $pp$ collisions.
\subsection{\bf Multiplicity dependent $b_{corr}(\Sigma p_{\rm T}$)}

\begin{figure}[hbt!]
\centering
\includegraphics[scale=0.30,keepaspectratio]{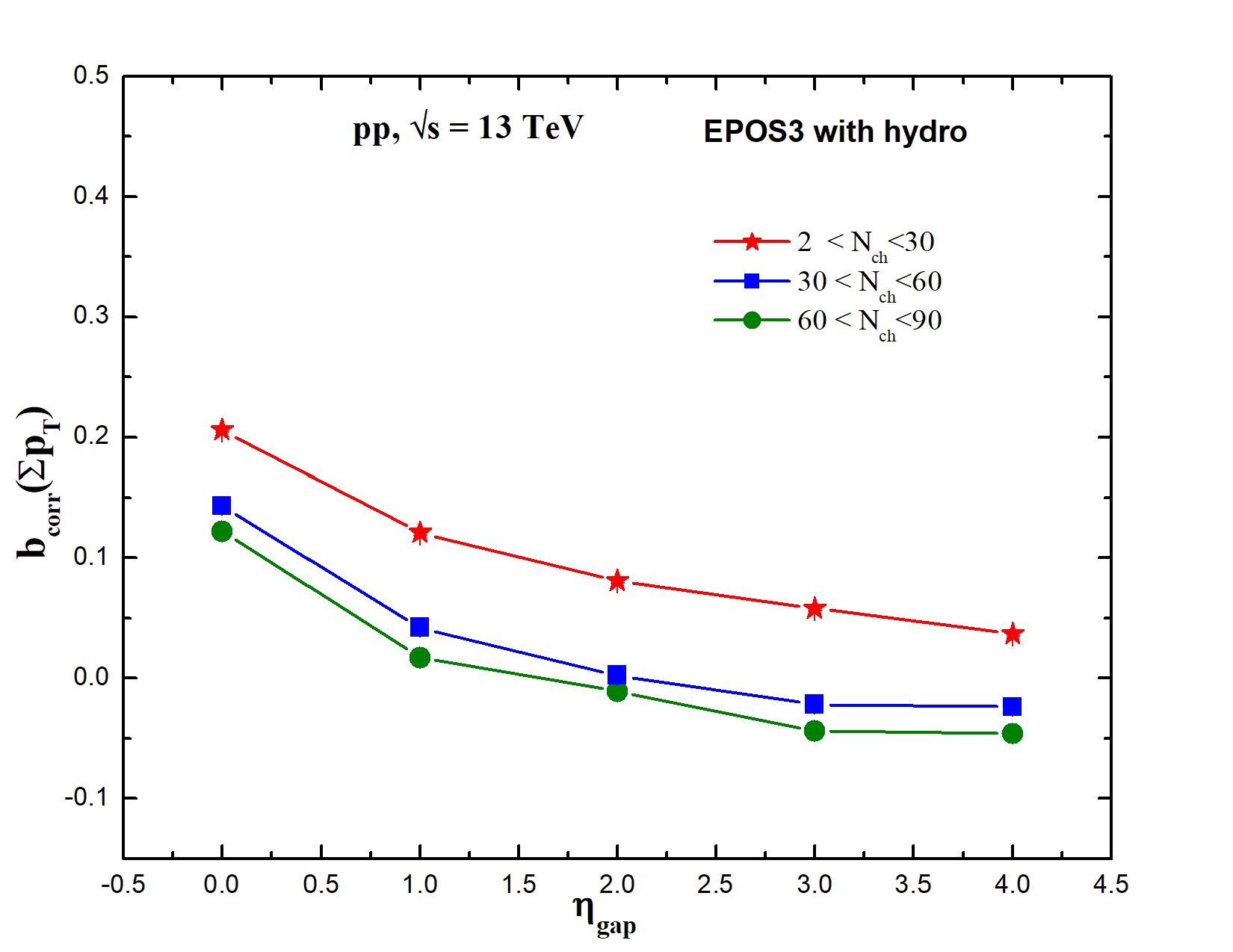}
\caption{Forward-backward summed-$p_{\rm T}$ correlations as a function of $\eta_{gap}$ for window width $\delta\eta$ = 0.5 
in different multiplicity ranges for EPOS3 simulated $pp$ events at $\sqrt{s}$ = 13 TeV.}
\label{fig11}
\end{figure}
\begin{figure}[hbt!]
\centering
\includegraphics[scale=0.30,keepaspectratio]{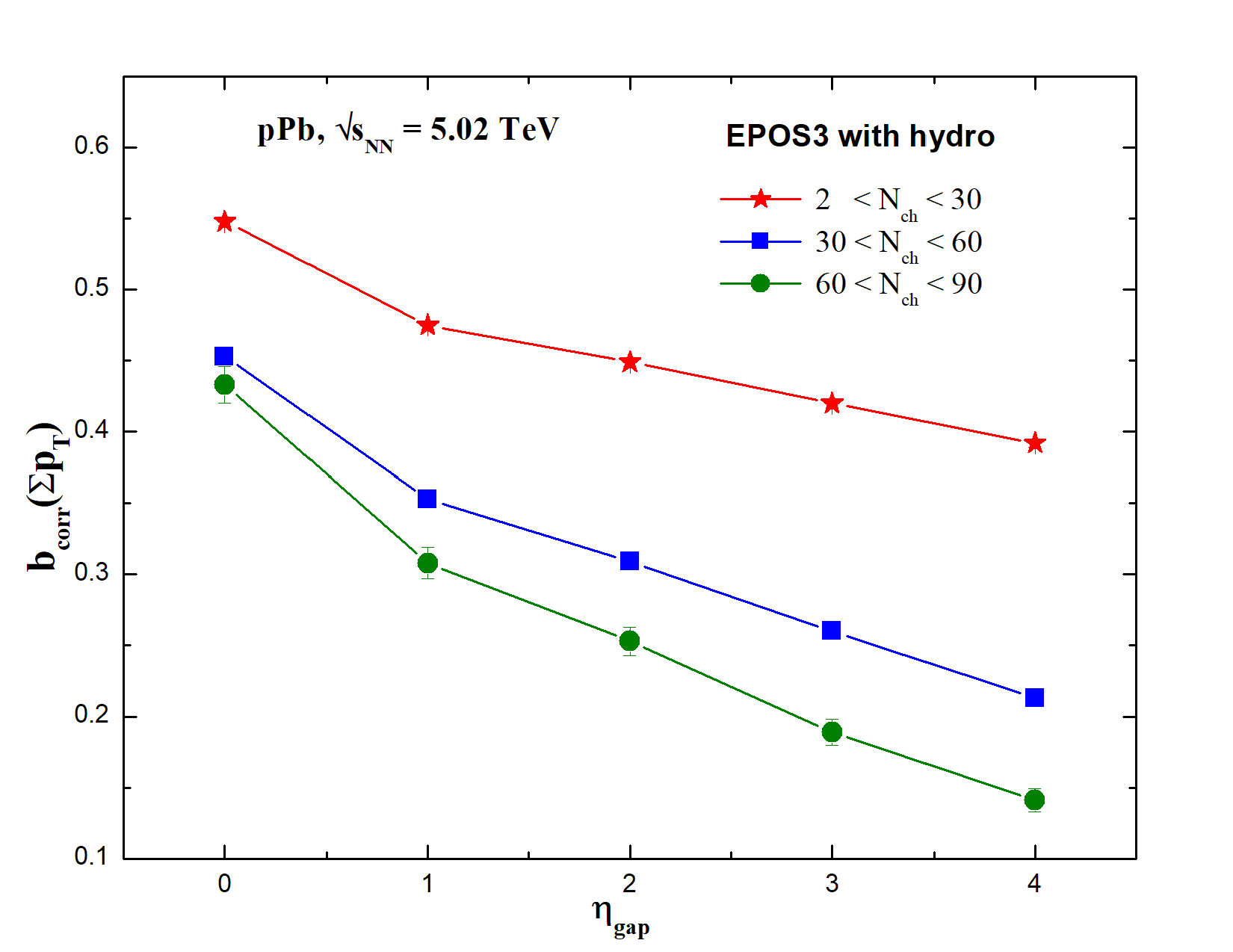}
\caption{Forward-backward summed-$p_{\rm T}$ correlations as a function of $\eta_{gap}$ for window width $\delta\eta$ = 0.5 
in different multiplicity ranges for EPOS3 simulated $pPb$ events at $\sqrt{s_{\rm NN}}$ = 5.02 TeV.}
\label{fig12}
\end{figure}

In addition to the study using minimum-bias EPOS3 events, we have exploited a multiplicity-dependent summed-$p_{\rm T}$ FB correlations study 
as well. Fig.~\ref{fig11} and Fig.~\ref{fig12} show FB momentum correlation as a function of $\eta_{gap}$ for $\delta\eta$ = 0.5 in three multiplicity 
ranges estimated following ATLAS kinematics~\cite{JM16} using EPOS3 simulated $pp$ events at $\sqrt{s}$ = 13 TeV and $pPb$ events at $
\sqrt{s_{\rm NN}}$ = 5.02 TeV. Red, blue and green points correspond to non-overlapping multiplicity regions; Low- (2 $< \rm N_{ch} <$ 30), Mid- (30 $< \rm N_{ch} <$  60) and High-multiplicity (60 $< \rm N_{ch} <$  90) regions respectively. We have kept the multiplicity ranges same for both $pp$ and $pPb$ events for better understanding.
We can see the similar decrease of correlation strength with increasing $\eta_{gap}$ and for a fixed $\eta_{gap}$ value, $b_{corr}(\Sigma p_{\rm T})$ also decreases with increasing multiplicity which may be due to the fact that fusion of strings into core in high multiplicity EPOS3 events lowers the FB  correlation strength~\cite{intro14}.
%
%
\section{Summary and Conclusions}\label{sec6}
We have performed a rigorous study on FB multiplicity and momentum correlation in $pp$ and $pPb$ collisions at the centre-of-mass energies, $\sqrt{s}$ = 13 TeV  and $\sqrt{s_{\rm NN}}$ = 5.02 TeV respectively at the LHC using EPOS3 simulated events with all charged particles and particles from core \& corona. We have investigated the behaviour of FB correlation strengths on gap between two pseudorapidity windows ($\eta_{gap}$), window width ($\delta\eta$), minimum transverse momentum ($p_{Tmin}$) and different multiplicity classes. Many LHC findings confirm a strong analogy between the small collision systems, $pp$ and $pPb$ particularly in terms of particles correlations and fluctuations~\cite{intro16, intro17, CMS2partC}. We have also noticed here that general trends of both FB correlation strengths are similar in case of both EPOS3 simulated with hydro $pp$ and $pPb$ collisions irrespective of energy difference.
We have found following results:
\begin{itemize}
    \item The linear relationship between $\langle N_{b}\rangle_{N_f}$ and $N_f$  is verified for EPOS3 generated both $pp$ and $pPb$ events and is more steeper for the $pPb$ events than $pp$ events.
    \item Our model based study fairly describes two general features of both FB correlation coefficients in two different collision systems ($pp$ and $pPb$) that it decreases slowly with the increase of gap between two selected $\eta$-windows irrespective of window widths and increases non-linearly with window width for a fixed separation between $\eta$-windows. 
    \item Rapid decrease in the values of both $b_{corr}(mult)$ and $b_{corr}(\sum{ p_T})$ with the small increase of minimum transverse momentum $p_{Tmin}$ has been observed for both $pp$ and $pPb$ events.
    \item Multiplicity dependent summed-$p_T$ correlation study also reveals that with the increase of multiplicity the value of $b_{corr}(\sum p_T)$ decreases at a fixed $\eta_{gap}$ for both $pp$ and $pPb$ events.
\end{itemize}
    
All these facts resemble our previous assessment of FB correlations in $pp$ collisions at three comparatively lower centre-of-mass energies, $\sqrt{s}$ = 0.9, 2.76 \& 7 TeV~\cite{intro14}. The observed larger FB multiplicity and momentum correlation strength in $pPb$ collisions w.r.t $pp$ collisions could be due to the fact that the initial asymmetry of the $pPb$ collisions and the large system size w.r.t the $pp$ collisions may enhance the event-by-event fluctuations which in turn may increase the FB correlation.\\
 
The most interesting result from our present study is the behaviour of $\delta\eta$-weighted average of FB multiplicity and momentum correlation strengths as a function of centre-of-mass energy ($\sqrt{s}$ = 0.9, 2.76, 7 \& 13 TeV) using EPOS3 simulated $pp$ events. The increase of both the correlation strengths ($b_{corr}(mult)$ \& $b_{corr}(\Sigma p_{\rm T})$) with $\sqrt{s}$ is clearly visible and interestingly we observed that it tends to saturate at very high energy. We have incorporated different theoretical model studies to compare our results and the correlation strengths are found to be following the similar trend as of EPOS3 simulated events. To investigate the possible reason behind such an interesting observation, we have calculated the correlation strengths for the EPOS3 generated events with all charged particles and particles from core \& corona. We found that for the particles from corona, the $\delta\eta$-weighted average of FB correlations do not show any saturation trend whereas for particles form core are perfectly exhibited the trend. We may infer that it may be due to the dominance of gluon-saturation effect at such higher centre-of-mass energies. 

Our analyses uncover the fact that the FB multiplicity and momentum correlation as a function of $\eta_{gap}$, $\delta\eta$, $p_{Tmin}$ and different multiplicity classes in EPOS3 simulated $pPb$ events qualitatively resemble the outcome of EPOS3 simulated $pp$ events though the values of correlation coefficients are higher for $pPb$ events than those for $pp$ events. Overall, we may conclude that the systematic study on FB correlations in different dimensions using the hybrid Monte-Carlo model EPOS3~\cite{epos1} adds more valuable information to understand the existing experimental results as well as encourages more experimental measurements at higher centre-of-mass energies and in different collision systems.
\section*{Acknowledgements}
The authors are thankful to Dr. Klaus Warner for providing us with the EPOS3 model.
The authors are thankful to the members of the grid computing team of VECC and cluster computing team of dept. of Physics, Jadavpur University 
for providing uninterrupted facility for event generation and analyses. We also gratefully acknowledge the financial help from the DST-GOI under the scheme ``Mega facilities in basic science research" [Sanction Order No.  SR/MF/PS-02/2021-Jadavpur (E-37128) dated 31.12.2021]. One of the authors (JM) acknowledges DST-INDIA for providing fellowship under INSPIRE Scheme.  
%

%
%
\end{document}